# Intelligent detect for substation insulator defects based on CenterMask

**Bo Ye\*, Feng Li, Mingxuan Li, Peipei Yan, Huiting Yang, Lihua Wang**

State Grid Xinjiang Electric Power Research Institute, Urumqi, China

**\* Correspondence:**
Bo Ye
bby90031@163.com

## Abstract

With the development of intelligent operation and maintenance of substations, the daily inspection of substations needs to process massive video and image data. This puts forward higher requirements on the processing speed and accuracy of defect detection. Based on the end-to-end learning paradigm, this paper proposes an intelligent detection method for substation insulator defects based on CenterMask. First, the backbone network VoVNet is improved according to the residual connection and eSE module, which effectively solves the problems of deep network saturation and gradient information loss. On this basis, an insulator mask generation method based on a spatial attention-directed mechanism is proposed. Insulators with complex image backgrounds are accurately segmented. Then, three strategies of pixel-wise regression prediction, multi-scale features and centerness are introduced. The anchor-free single-stage target detector accurately locates the defect points of insulators. Finally, an example analysis is carried out with the substation inspection image of a power supply company in a certain area to verify the effectiveness and robustness of the proposed method.

## 1    Introduction

In China, the "State Grid Corporation Artificial Intelligence Technology Application 2021 Work Plan" has made clear instructions. To build and operate two levels of AI "two libraries and one platform", and form a number of high-precision and high-value power-specific models. As an important and indispensable part of the power system, the safe and stable operation of substations is the basis for building a strong smart grid (Han et al., 2021, Li et al. 2022a). Today, rising renewable energy uncertainty (Li et al. 2022b, Li et al. 2022c) and concerns about cyber security and privacy issues (Li et al. 2019, Li et al. 2022d) pose new challenges for substation operations. Therefore, there are strict standards for operation and maintenance (Q&M) of substation equipment (Huang et al., 2017).

Insulators are key equipment in substations. Common types include: pillar insulators, bushing insulators and suspension insulators. Its operational performance is directly related to the stable and continuous operation of the regional power grid. However, insulator fouling and breakage are the main factors affecting its operational performance (He et al., 2017). In order to meet the needs of Q&M, the substation is equipped with inspection robots and deployed a large number of high-definition cameras (Guan et al., 2021). The daily Q&M of insulators in the station needs to process massive amounts of video and image data. How to use data to efficiently analyze insulator defects is not only the focus of Q&M work, but also the difficulty in defect detection (Wang et al., 2017).

Currently, there have already been a significant number of investigations on the defects detection of insulators in substations, as artificial intelligence has been successfully applied to classification problems in engineering (Shi et al. 2008, Shi et al. 2009). In (Zhai et al., 2018), based on machine vision, the feature extraction of the insulator image is completed through the Vision module of LabVIEW, and the preliminary identification and positioning of the insulator is realized. The surface of substations insulators is very susceptible to the influence of wet weather, resulting in contamination and flashover accidents in substation facilities. Ref. (Tang et al., 2020) designed an insulator cleaning robot based on binocular vision, using the YOLOv4-tiny deep learning network algorithm to identify and detect pillar insulators and flange targets. By establishing an electrothermal conversion model for insulator strings, a method for identifying deteriorating porcelain insulators in substations based on infrared thermal imaging technology is proposed. The work in (Zhang et al., 2020) obtains a degree of practicality. In addition, Ref. (Gao et al., 2021) in order to improve the fault diagnosis capability of substation equipment insulators. A method for identifying breakage of substation equipment insulators based on intelligent image information fusion and edge contour segmentation detection is proposed. In (Usamentiaga et al., 2018) studied multi-source image information fusion detection and multi-source image fusion reconstruction method of insulator contamination state. However, this work extracts the surface color features, temperature rise features, and discharge intensity features of insulators from visible light images, infrared images, and ultraviolet images, respectively, which makes the detection method difficult to apply online (Jin et al., 2018).

In summary, the application of computer vision and deep learning methods to realize defect detection of substation insulators has become a research hotspot. However, the existing methods are not ideal for the detection of massive high-definition image data in substations, and there is a common problem of low real-time detection of insulator defects. Therefore, based on an end-to-end learning paradigm, this work adopts a one-stage instance segmentation algorithm, and an end-to-end agile detection method for substation insulator defects is proposed.

The main novelties of this work are:

(i) The end-to-end design idea revolutionizes the staged and step-by-step segmentation and detection methods in traditional insulator detection research. This strategy effectively reduces the impact of step-by-step error iteration on model performance.

(ii) Insulator mask generation and defect detection can be performed in parallel. This method greatly improves the fault identification efficiency of the equipment Q&M in the substation.

(iii) The one-stage instance segmentation algorithm has been successfully applied in the detection of insulator defect points. By introducing pixel-by-pixel regression prediction, multi-scale features and centerness three strategies to accurately output the bounding box label of the defect point location.

This paper is organized as follows: Section 2 briefly describes the architecture for agile detection of insulator defect points in substations. The method of insulator mask generation and defect point detection is described in detail in Section 3. Section 4 verifies the effectiveness of the proposed method based on the built experimental environment. Finally, conclusions and outlook are presented in Section 5.

## 2  The architecture of insulator defect intelligent detection method

Due to the close connection between the equipment in the substation, the insulator image taken by the inspection robot is mixed with more other transformer equipment (Wang et al., 2019; Kou et al., 2022). This makes it necessary to step through the images during insulator defect detection. However, the result of step-by-step processing will certainly increase the detection time and reduce the accuracy of detection (Zhao et al., 2019; Chen et al., 2019). At the same time, the characteristics of the massive amount of inspection images are different to ignore. Therefore, this study comprehensively measures the Q&M requirements of substation insulators and the training complexity of massive inspection images. A defect detection method based on CenterMask is designed. The overall implementation plan of this work is shown in **Figure 1**. Based on the one-stage target detection algorithm, a parallelized image segmentation technique is incorporated to obtain the insulator mask (Yuan et al., 2021). At the same time, the separate extraction of masks for insulator defect points is also incorporated into the training. By counting the insulator defect detection task points, the main tasks are: insulator detection, insulator mask segmentation, insulator defect point detection, and insulator defect point location.

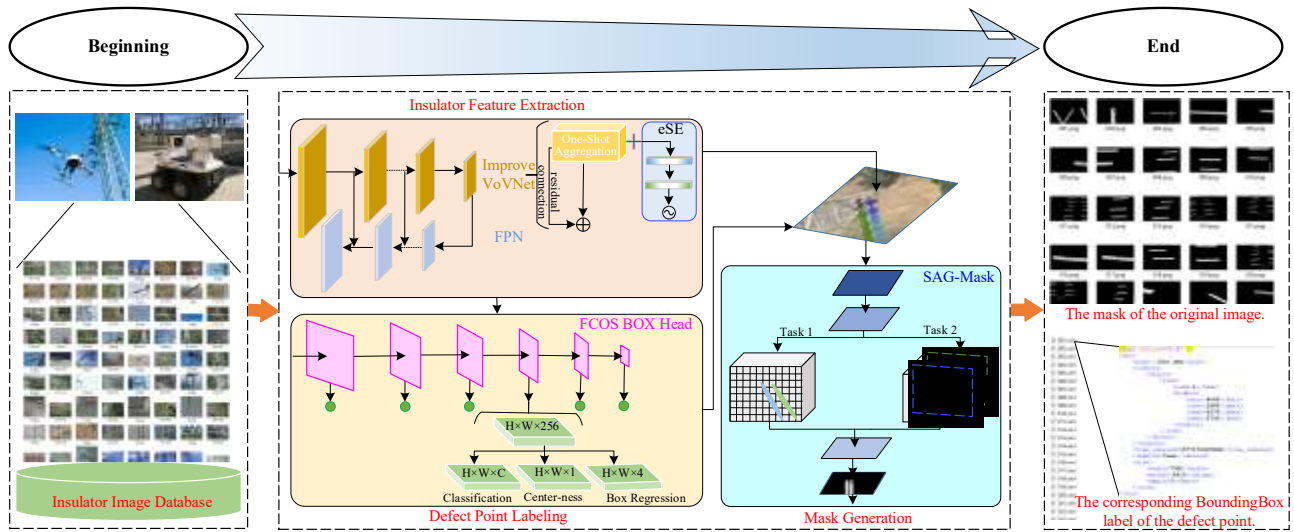

**Figure 1** The overall framework of the intelligent detection method for insulator defects

As shown in Figure 1, this end-to-end work can be decomposed into three parts. First, based on the accumulated original inspection images, the initial image database is constructed. Then, the defect detection model structure is designed. This part is also the core of the whole method. 1)The backbone network adopts improved VoVNet to achieve efficient extraction of target features. 2)The spatial attention-guided mask branch (SAG-Mask) generates target masks. 3)Using fully convolutional one-stage (FCOS) to detect defect points of insulators. Finally, the end realizes the output of the insulator mask and the bounding box label of the defect point.

## 3  Agile Detection Method for Substation Insulator defects

CenterMask is a newer instance segmentation algorithm proposed by Youngwan Lee and Jongyoul Park of the Korea Institute of Electronics and Communications in 2019 (Lee et al., 2020). This algorithm adds the SAG-Mask module invented by scholars to the famous one-stage anchor-free target detection algorithm FCOS, which makes its target segmentation accuracy and speed superior to traditional algorithms. The two scholars made the algorithm open source for scholars in the field of image and vision to jointly explore the application research of the algorithm. The application of this algorithm in substation insulator defect detection further improves the intelligence and accuracy of insulator defect detection.

## 3.1 Insulator feature extraction based on improved VoVNet

The inspection robot in the smart substation will collect images of insulators from different angles during the operation. The background of the insulator in the image is complex and diverse, which has serious noise interference to the accurate detection of insulator defect points (Tao et al., 2018; Kou et al., 2020). Therefore, our first task is to achieve accurate extraction of insulator target features. It is the backbone network of the model that accomplishes this task in the study.

At present, various target detection models rely on different backbone networks (Du et al., 2021). The core module used by the mainstream target detection model DenseNet is the dense block. All previous layers are aggregated through dense connections, resulting in a linear increase in the number of input pipelines for each subsequent layer. This strategy makes the memory access cost and energy consumption extremely high, and the computing speed is seriously hindered (Xu et al. 2020). When the input is a higher-quality insulator image, object extraction tends to consume more memory and inference time. This problem makes the detection effect of insulator defects far from meeting the standard of massive high-definition insulator images in substations. The CenterMask backbone network proposed by the research utilizes an improved VoVNet. The short board of feature redundancy caused by dense connections in DenseNet is effectively solved, and the core module of one-shot Aggregation (OSA) is adopted. In the strategy, all the previous layers are unified and aggregated at the end.

The improved VoVNet addresses the performance saturation problem and the information loss problem in standard VoVNet, respectively. 1)By adjusting the input of the OSA module and adding it to the output, the residual link of ResNet is introduced to ensure that VoVNet can train a deeper network. It can effectively process insulator images with high pixel ratio. 2)The eSE module of SENet is added to the last feature layer of VoVNet, and the two fully connected layers used in the original SE module are directly replaced by one, which effectively reduces the loss of channel information. This method further enhances the feature extraction capability of the insulator target. The improved connection structure of OSA modules in VoVNet according to the idea is shown in the insulator feature extraction area in **Figure 1**. The performance of the target detection models based on the improved VoVNet surpasses the models based on other target detection algorithms. The algorithm greatly improves the GPU computing efficiency and meets the needs of high-speed processing for intelligent identification of hidden dangers of insulators.

## 3.2 Insulator mask extraction based on SAG-Mask

Insulator image datasets mostly suffer from the problems of complex backgrounds, multiple strings in a single image and insulator overlap (Wang et al., 2020b). Section 3.1 Insulator feature extraction realizes insulator target detection. The position of the insulator in the image is accurately framed, and only completing this link cannot quickly detect the defect points of the insulator. Therefore, it is necessary to further realize the precise segmentation of insulator beads on this basis. Adding the SAG-Mask branch to the target result of FCOS detection in the CenterMask algorithm satisfactorily completes this task.

The most famous ones in instance segmentation are Mask RCNN and its related improved algorithms, as well as the YOLACT algorithm which focuses on speed (Wang et al., 2020b).

However, a spatial attention-oriented mechanism is proposed in CenterMask. The mechanism is able to utilize the spatial attention map of the input image to predict segmentation masks for each boxed instance. The added SAG-Mask branch can be performed in parallel with the insulator fault point identification of the target detector, which greatly improves the identification efficiency of intelligent hidden dangers of insulators.

In this paper, the instance segmentation of insulators is decomposed into two simple and parallel subtasks, as shown in the Mask generation area in **Figure 1**. Task 1 achieves the goal of constraining the local area of each insulator and naturally distinguishing the instance. A rough shape prediction is made around the center point of each insulator. Task 2 achieves precise segmentation of insulators while retaining the spatial position of precise segmentation. The branch is predicted using the insulator saliency pixels in the entire inspection image. The core idea is to focus on specific block features in the feature map by using an attention mechanism. Finally, a mask for each insulator's corresponding image position is constructed by multiplying the outputs of the two parallel branches.

The mathematical process of instance segmentation process can be described as: the input insulator feature map is marked as $X_i \in R^{C*H*W}$. It is pooled using max-pooling and average-pooling. The pooled features are respectively $F_{max\_pooling}, F_{average\_pooling} \in R^{1*H*W}$ and subsequently passed through a $3 \times 3$ convolutional neural network. The specific mathematical functions are as follows:

$$A_{sag}(X_i) = \sigma(F_{3*3}(P_{max\_pooling} \cdot P_{average\_pooling})) \qquad (1)$$

Then, saliency pixels are used in task 2 to enhance the input features of the original insulator image. As shown in equation (2).

$$X_{sag} = A_{sag}(X_i) \otimes X_i \qquad (2)$$

### 3.3 Identification and location of insulator defects based on FCOS

The goal of intelligent hidden danger identification of insulators is to quickly locate various common faults of insulators through intelligent algorithms, such as bead damage, excessive pollution, aging cracks, etc (Wang et al., 2020b). The accurate extraction of the insulator mask can only ensure that the location of the insulator is identified from the inspection image, and the insulator is accurately segmented. However, specific fault points require accurate location detection of the corresponding model. This work uses the anchor-free FCOS algorithm to complete the task in the target detection part of the overall CenterMask structure. FCOS is a fully convolutional one-stage object detection algorithm (Zhang et al., 2022). It solves the object detection problem in a pixel-by-pixel prediction manner. At the same time, in the detection process, the anchor-free box of FOCS is a huge advantage. The hyperparameters associated with anchor boxes and all the complex computations associated with anchor boxes are effectively avoided. The efficiency of insulator defect identification and location has been greatly improved. The output of the entire defect identification and localization results includes three aspects: pixel-by-pixel regression prediction, multi-scale features and center-ness.

The specific process framework is shown in the defect point labeling area in **Figure 1**. The first output of the 3 output layers is the classification branch. H*W represents the size of the feature. C represents the number of categories. A position in the input inspection image can be mapped to a position on the feature, and the mapping function is:

$$(\left\lfloor\frac{r}{2}\right\rfloor + p\hat{r}, \left\lfloor\frac{r}{2}\right\rfloor + qr) \tag{3}$$

where the coordinates $(p,q)$ denote a specific location in the inspection image. $r$ denotes the scaling between the feature map and the original inspection image. This function can represent the relationship between the position of the point on the feature map and the position of the point on the output image. This lays the groundwork for computing the classification and regression objectives for each point on the feature map.

The second output is the center-ness policy. The FCOS algorithm is applied to achieve high efficiency of one-stage anchor-free box calculation and high recall rate of pixel-by-pixel regression strategy. It also brings many low-quality prediction bounding boxes with too large deviation from the center point. By introducing this strategy, the distance between each point and the target center can be calculated, thereby suppressing some predicted bounding boxes that are far away from the target center. But this strategy does not introduce any excessive hyperparameters.

The centerness strategy adds a branch to each prediction layer of the feature pyramid. And this branch is parallel to the classification branch, which is equivalent to adding a loss function to the network. During the training process of the image dataset, the loss function constrains the predicted bounding box to be as close to the center point as possible, so that the auxiliary non-maximum suppression (NMS) can filter out the low-quality bounding box prediction. The loss function is as follows:

$$Loss^* = \sqrt{\frac{\min(d_{left}^*, d_{right}^*)}{\max(d_{left}^*, d_{right}^*)} \times \frac{\min(d_{top}^*, d_{bottom}^*)}{\max(d_{top}^*, d_{bottom}^*)}} \tag{4}$$

where the distances from the center point to the left, right, top and bottom sides of the prediction bounding box are represented by $d_{left}$, $d_{right}$, $d_{top}$ and $d_{bottom}$, respectively. $Loss^*$ is the value of the loss function, and the more it tends to zero, the better the prediction can be constrained.

The third output is the return branch. When performing the target frame regression on all the points in the target frame of defect points in the inspection image, the distance to each edge is used as the measurement standard. This part is the main difference between the target detection algorithm without anchor frame and the target detection algorithm based on anchor frame, where 4 in $H \times W \times 4$ represents 4 values related to regression. Calculated as follows:

$$\begin{aligned} d_{left}^* &= m - m_0^{(i)}, d_{top}^* = n - n_0^{(i)} \\ d_{right}^* &= m_0^{(i)} - m, d_{bottom}^* = n_0^{(i)} - n \end{aligned} \tag{5}$$

## 4  Case Study

In order to verify the effectiveness and robustness of the insulator defect agile detection method proposed in this paper. Taking the substation inspection image of a power supply company in a certain area as an example, an example analysis is carried out. There are two main aspects to build the experimental environment in DELL graphics workstation. Hardware: Intel(R) Xeon(R) Gold5118 CPU @ 2.30GHz, NVIDIA Quadro P5000, 128G RAM; Software: Python 3.7.6, CUDA 10.1,

PyTorch 1.4.0, Detectron2 0.1.1, CV2 4.2.0. The model output results are evaluated by a number of index combinations to further realize the optimization of the defect point identification model.

## 4.1  Insulator mask generation and defect detection

The insulator images captured by the inspection are divided into training set and test set. In the above experimental environment, the analysis of training set samples, the selection and optimization of the model backbone network are carried out. This model was trained over 200,000 iterations. During the training process, each part of the network loss is visualized to achieve accurate tuning of model parameters. The complete flow of insulator mask generation is shown in **Figure 2**.

First, the training set as shown in **Figure 2(a)** is manually annotated. It is used to assist the initial learning of the model. **Figure 2(b)** shows the manual annotation of the original inspection image using the labeling tool Labelme. Insulators are marked with the "Insulator" label, and insulator defect points are marked with the "Insulator error" label. Then, the CenterMask network model is iteratively learned and adjusted using the effectively labeled training set. This process takes a long time. Finally, the trained insulator defect detection model is applied. Input test set for model insulator mask extraction branch validity verification. The extracted rendering is shown in **Figure 2(c)**.

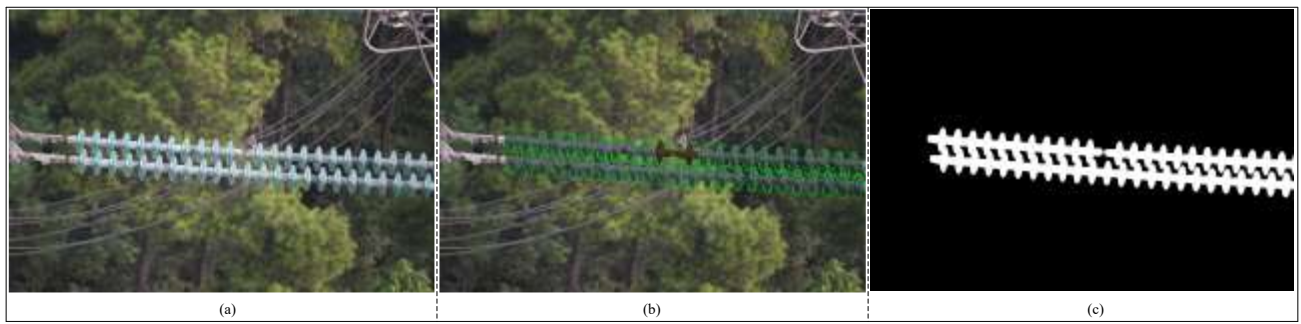

(a)　　　　　　　　　　　　　　(b)　　　　　　　　　　　　　　(c)

**Figure 2** The insulator mask generation process: (a) Original image of insulator beading. (b) Annotated drawing of the insulator. (c) Insulator mask extraction.

As mentioned in Section 2, defect point identification and mask extraction in the insulator defect detection model are two independent branches, and the two branches can be executed in parallel. The identification of defect points does not have to wait for mask generation or be disturbed by mask generation results. Therefore, the defect point identification and positioning is more flexible and the effect is better. By adopting the same multi-scale training strategy as insulators, the model parameters are set for random scaling within the range of input scales. The insulator explosion point training dataset is enriched according to the maximum and minimum input size constraints. When outputting the defect point xml file, the visual output display of single image insulator defect point identification and identification accuracy is realized. For the test set model insulator defect point detection visualization diagram and the corresponding positioning label stored in the xml file effect shown in **Figure 3**.

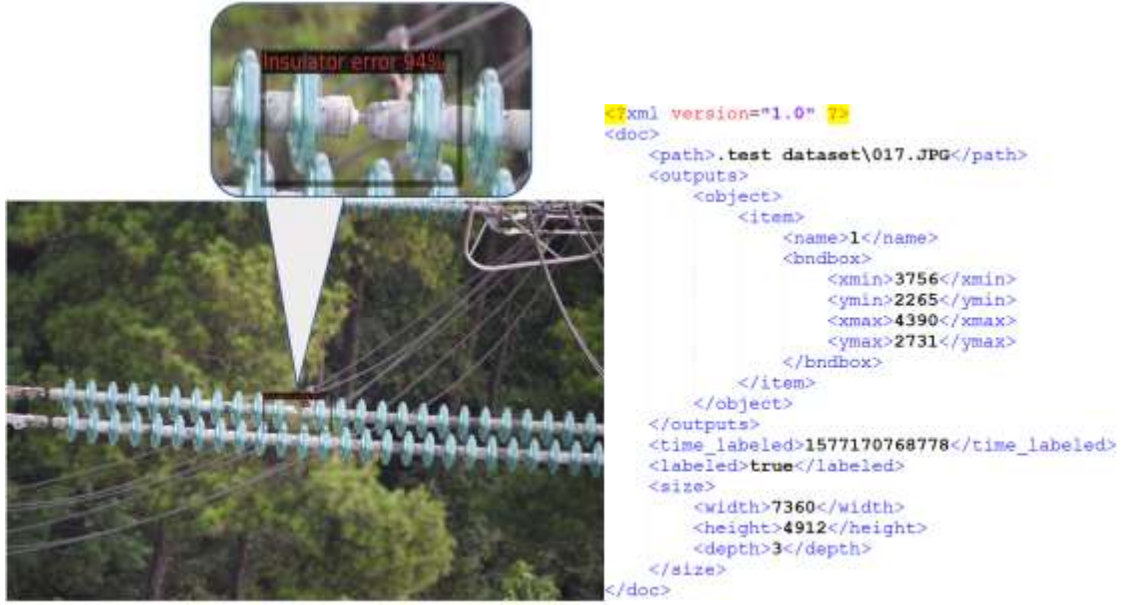

**Figure 3** The insulator defect point detection and positioning.

### 4.2 Example index evaluation

For the model test results, the average precision (AP) and average recall (AR) evaluation metrics are set. Under the combination of different IOUs, areas under inspection (AREA) and number of objects under inspection (MaxDets), model performance analysis is performed on all test results. The specific evaluation results are shown in **Table 1**. Where IOU represents the degree of overlap between the generated candidate frame and the original marked frame. It is the ratio of intersection and union, and the calculation equation is as follows:

$$\text{IOU} = \frac{\text{Area of Overlap}}{\text{Area of Union}} \tag{6}$$

**Table 1** The index values of different assessment methods

| Assessment method | IOU | AREA | MaxDets | Value1 | Value2 |
|---|---|---|---|---|---|
| AP | 0.50-0.95 | all | 100 | 0.921 | 0.881 |
| $AP_{50}$ | >0.50 | all | 100 | 0.934 | 0.926 |
| $AP_{75}$ | >0.75 | all | 100 | 0.854 | 0.873 |
| $AP_s$ | 0.50-0.95 | small | 100 | 0.845 | 0.694 |
| $AP_m$ | 0.50-0.95 | medium | 100 | 0.896 | 0.811 |
| $AP_l$ | 0.50-0.95 | large | 100 | 0.912 | 0.976 |
| $AR_{m1}$ | 0.50-0.95 | all | 1 | 0.632 | 0.694 |
| $AR_{m10}$ | 0.50-0.95 | all | 10 | 0.726 | 0.789 |
| $AR_{m100}$ | 0.50-0.95 | all | 100 | 0.791 | 0.804 |
| $AR_s$ | 0.50-0.95 | small | 100 | 0.751 | 0.792 |
| $AR_m$ | 0.50-0.95 | medium | 100 | 0.887 | 0.834 |
| $AR_l$ | 0.50-0.95 | large | 100 | 0.898 | 0.942 |

In **Table 1**, AP represents IOU ranging from 0.5~0.95. $AP_{50}$ represents the IOU range of 0.5~1.0. $AP_{75}$ represents the IOU range of 0.75~1.0. $AP_s$ represents the AP measurement of the target box whose pixel area is smaller than $32^2$. $AP_m$ represents the AP measurement of the target box whose pixel area is between $32^2 \sim 96^2$. $AP_l$ represents the AP measurement of the target box whose pixel area is larger than $96^2$. Value1 represents the AP/AR generated by the insulator mask under different evaluation methods. Value2 represents the index result value of defect point detection in different situations.

Analysis of **Table 1** shows that when 0.5 is used as the IOU threshold for segmentation, the model mask generation effect is the best. The accuracy of the model is 93.4%, indicating that the model satisfies the extraction of insulators in most inspection image scenarios. However, the AP and AR of target boxes with a pixel area smaller than are significantly lower, indicating that the model still needs to be improved in the details of mask generation. The performance of the model in defect point detection is further analyzed, and the overall Q&M requirements of the substation are met, and the detection effect is good. In the detection of objects with a pixel area greater than $96^2$, AP and AR are 97.6% and 94.2%, respectively. The model performs best. It can be seen that the model has a high detection accuracy rate for large insulator defect points, and has a good recall rate.

# 5      Conclusion

This work takes the intelligent Q&M of the equipment in the substation as the background, and comprehensively analyzes the characteristics of the massive images of inspection insulators and the complex background of defect point detection. An intelligent detection method for substation insulator defects based on CenterMask is proposed. The proposal of this technology improves the efficiency and accuracy of insulator defect detection to a certain extent. The results of the example analysis verify the following main conclusions:

1) An end-to-end insulator defect point detection architecture is designed. The entire learning process does not perform artificial sub-problem division, but the deep learning model directly learns the mapping from the original input to the desired output.

2) It is only necessary to input the insulator field image captured by the inspection robot into the defect detection model. The three key tasks of target feature extraction, mask segmentation and defect point target detection of insulators can be realized at one time.

3) A new agile detection method for insulator defects in substations is proposed. In the manual inspection mode, the waste of human resources and the potential safety hazards caused by the operation of the power grid are effectively reduced.

The transfer application research of this method will be the next exploration direction. It will greatly promote the efficiency and intelligence of substation equipment Q&M, and has important research significance and engineering application value.

# 6      Data Availability Statement

The original contributions presented in the study are included in the article/supplementary material, further inquiries can be directed to the corresponding author.

# 7      Author Contributions

BY: designed this study. FL: contributed to the architecture of insulator defect intelligent detection method. ML: contributed to the insulator feature extraction based on improved VoVNet. PY: contributed to the insulator mask extraction based on SAG-Mask. HY: performed the identification and location of insulator defects based on FCOS. LW: collected and cleansed the data. All authors contributed to the writing of the article and all agreed to the submitted version of the article.

## 8 Conflict of Interest

The authors declare that the research was conducted in the absence of any commercial or financial relationships that could be construed as a potential conflict of interest.

## 9 References

Chen, J., Xu, X., and Dang, H. (2019). Fault detection of insulators using second-order fully convolutional network model. Mathematical Problems in Engineering, 2019. doi: 10.1155/2019/6397905.

Du, Y., Du, L., and Li, L. (2021). An SAR Target Detector Based on Gradient Harmonized Mechanism and Attention Mechanism. IEEE Geoscience and Remote Sensing Letters, 19, 1-5. doi: 10.1109/LGRS.2021.3103378.

Gao, Z., Yang, G., Li, E., and Liang, Z. (2021). Novel Feature Fusion Module-Based Detector for Small Insulator Defect Detection. IEEE Sensors Journal, 21(15), 16807-16814. doi: 10.1109/JSEN.2021.3073422.

Guan, X., Gao, W., Peng, H., Shu, N., and Gao, D. W. (2021). Image-Based Incipient Fault Classification of Electrical Substation Equipment by Transfer Learning of Deep Convolutional Neural Network. IEEE Canadian Journal of Electrical and Computer Engineering, 45(1), 1-8. doi: 10.1109/ICJECE.2021.3109293.

Han, S., Yang, F., Jiang, H., Yang, G., Zhang, N., and Wang, D. (2021). A Smart Thermography Camera and Application in the Diagnosis of Electrical Equipment. IEEE Transactions on Instrumentation and Measurement, 70, 1-8. doi: 10.1109/TIM.2021.3094235.

He, J., and Gorur, R. S. (2017). Flashover of insulators in a wet environment. IEEE Transactions on Dielectrics and Electrical Insulation, 24(2), 1038-1044. doi: 10.1109/TDEI.2017.005795.

Huang, Q., Jing, S., Li, J., Cai, D., Wu, J., and Zhen, W. (2016). Smart substation: State of the art and future development. IEEE Transactions on Power Delivery, 32(2), 1098-1105. doi: 10.1109/TPWRD.2016.2598572.

Jin, L., Tian, Z., Ai, J., Zhang, Y., and Gao, K. (2018). Condition evaluation of the contaminated insulators by visible light images assisted with infrared information. IEEE Transactions on Instrumentation and Measurement, 67(6), 1349-1358. doi: 10.1109/TIM.2018.2794938.

Lee, Y., and Park, J. (2020). Centermask: Real-time anchor-free instance segmentation. In Proceedings of the IEEE/CVF conference on computer vision and pattern recognition (pp. 13906-13915). doi: 10.1109/CVPR42600.2020.01392.

Kou, L., Li, Y., Zhang, F., Gong, X., Hu, Y., Yuan, Q., and Ke, W. (2022). Review on Monitoring, Operation and Maintenance of Smart Offshore Wind Farms. Sensors, 22(8), 2822. doi: 10.3390/s22082822.

Kou, L., Liu, C., Cai, G. W., Zhou, J. N., Yuan, Q. D., & Pang, S. M. (2020). Fault diagnosis for open-circuit faults in NPC inverter based on knowledge-driven and data-driven approaches. IET Power Electron, 13(6), 1236-1245. doi: 10.1049/iet-pel.2019.0835

Li, Y., Li, Z., and Chen, L. (2019). Dynamic state estimation of generators under cyber attacks. IEEE Access, 7, 125253-125267. DOI:10.1109/ACCESS.2019.2939055.

Li, Y., Zhang, M., and Chen, C. (2022a). A deep-learning intelligent system incorporating data augmentation for short-term voltage stability assessment of power systems. Applied Energy, 308, 118347. DOI:10.1016/j.apenergy.2021.118347.

Li, Y., Li, K., Yang, Z., et al. (2022b). Stochastic optimal scheduling of demand response-enabled microgrids with renewable generations: An analytical-heuristic approach. Journal of Cleaner Production, 330, 129840. DOI:10.1016/j.jclepro.2021.129840

Li, Y., Wang, B., Yang, Z., et al, (2022c). Hierarchical stochastic scheduling of multi-community integrated energy systems in uncertain environments via Stackelberg game. Applied Energy, 308: 118392. DOI:10.1016/j.apenergy.2021.118392

Li, Y., Li, J., and Wang, Y. (2022d). Privacy-preserving spatiotemporal scenario generation of renewable energies: A federated deep generative learning approach. IEEE Transactions on Industrial Informatics, 18(4), 2310-2320. DOI:10.1109/TII.2021.3098259

Shi, Z. B., Yu, T., Zhao, Q., et al. (2008). Comparison of algorithms for an electronic nose in identifying liquors. Journal of Bionic Engineering 5(3), 253-257. DOI:10.1016/S1672-6529(08)60032-3

Shi, Z. B., Li, Y., and Yu, T. (2009). Short-term load forecasting based on LS-SVM optimized by bacterial colony chemotaxis algorithm. In 2009 International Conference on Information and Multimedia Technology. 306-309. DOI:10.1109/ICIMT.2009.57

Tao, X., Zhang, D., Wang, Z., Liu, X., Zhang, H., and Xu, D. (2018). Detection of power line insulator defects using aerial images analyzed with convolutional neural networks. IEEE Transactions on Systems, Man, and Cybernetics: Systems, 50(4), 1486-1498. doi: 10.1109/TSMC.2018.2871750.

Tang, S., Zhou, P., Wang, X., Yu, Y., and Li, H. (2020). Design and experiment of dry-ice cleaning mechanical arm for insulators in substation. Applied Sciences, 10(7), 2461. doi: 10.3390/app10072461.

Usamentiaga, R., Fernandez, M. A., Villan, A. F., and Carus, J. L. (2018). Temperature monitoring for electrical substations using infrared thermography: architecture for Industrial Internet of Things. IEEE transactions on industrial informatics, 14(12), 5667-5677. doi: 10.1109/TII.2018.2868452.

Wang, H., Zhou, B., and Zhang, X. (2017). Research on the remote maintenance system architecture for the rapid development of smart substation in China. IEEE Transactions on Power Delivery, 33(4), 1845-1852. doi: 10.1109/TPWRD.2017.2757939.

Wang, H., and Meng, F. (2019). Research on power equipment recognition method based on image processing. EURASIP Journal on Image and Video Processing, 2019(1), 1-11. doi: 10.1186/s13640-019-0452-5.

Wang, B., Dong, M., Ren, M., Wu, Z., Guo, C., Zhuang, T., Pischler, O., and Xie, J. (2020a). Automatic fault diagnosis of infrared insulator images based on image instance segmentation and temperature analysis. IEEE Transactions on Instrumentation and Measurement, 69(8), 5345-5355. doi: 10.1109/TIM.2020.2965635.

Wang, S., Liu, Y., Qing, Y., Wang, C., Lan, T., and Yao, R. (2020b). Detection of insulator defects with improved resnest and region proposal network. IEEE Access, 8, 184841-184850. doi: 10.1109/ACCESS.2020.3029857.

Wen, Q., Luo, Z., Chen, R., Yang, Y., and Li, G. (2021). Deep learning approaches on defect detection in high resolution aerial images of insulators. Sensors, 21(4), 1033. doi: 10.3390/s21041033.

Yuan, Q., Zhang, Z., Pi, Y., Kou, L., and Zhang, F. (2021). Real-Time Closed-Loop Detection Method of vSLAM Based on a Dynamic Siamese Network. Sensors, 21(22), 7612. doi: 10.3390/s21227612.

Xu, D., and Wu, Y. (2020). Improved YOLO-V3 with DenseNet for multi-scale remote sensing target detection. Sensors, 20(15), 4276. doi: 10.3390/s20154276.

Zhai, Y., Chen, R., Yang, Q., Li, X., and Zhao, Z. (2018). Insulator fault detection based on spatial morphological features of aerial images. IEEE Access, 6, 35316-35326. doi: 10.1109/ACCESS.2018.2846293.

Zhao, Z., Zhen, Z., Zhang, L., Qi, Y., Kong, Y., and Zhang, K. (2019). Insulator detection method in inspection image based on improved faster R-CNN. Energies, 12(7), 1204. doi: 10.3390/en12071204.

Zhang, D., and Chen, S. (2020). Intelligent Recognition of Insulator Contamination Grade Based on the Deep Learning of Ultraviolet Discharge Image Information. Energies, 13(19), 5221. doi: 10.3390/en13195221.

Zhang, K., Qian, S., Zhou, J., Xie, C., Du, J., and Yin, T. (2022). ARFNet: adaptive receptive field network for detecting insulator self-explosion defects. Signal, Image and Video Processing, 1-9. doi: 10.1007/s11760-022-02186-3.